\documentclass[a4paper,twocolumn,english,aps,prb]{revtex4}
\usepackage[T1]{fontenc}
\usepackage[latin1]{inputenc}
\usepackage{amsmath}
\usepackage{babel}
\usepackage{amssymb}

\makeatletter

\providecommand{\LyX}{L\kern-.1667em\lower.25em\hbox{Y}\kern-.125emX\@}

\newcommand{\C}[1]{{}} % Comment
\def\ubar#1{\vtop{\ialign{##\crcr
      $\hfil\displaystyle{#1}\hfil$\crcr\noalign{\kern0pt\nointerlineskip}
      $\hfil\underline{\hbox to 4pt{}}\hfil$\crcr\noalign{\kern0pt}}}
}
\DeclareRobustCommand{\mat}[1]{\mathbf{#1}}
\newcommand{\bS}{\begin{subequations}}
\newcommand{\eS}{\end{subequations}}
\newcommand{\para}{\parallel}

\makeatother
\begin{document}
\C{

\newcommand{\mat}[1]{\underline{#1 }}

\newcommand{\para}{||}

}

\title{Anisotropy of ultra-thin ferromagnetic films and the spin reorientation
transition}

\author{K.~D.~Usadel}

\author{A.~Hucht}

\email{fred@thp.Uni-Duisburg.de}

\affiliation{Theoretische Physik, Gerhard-Mercator-Universität, D-47048 Duisburg, Germany}

\date{24th Feb. 2002}


\begin{abstract}
The influence of uniaxial anisotropy and the dipole interaction on the
direction of the magnetization of ultra-thin ferromagnetic films in the
ground-state is studied. The ground-state energy can be expressed in terms
of anisotropy constants which are calculated in detail as function of the
system parameters and the film thickness. In particular non-collinear spin
arrangements are taken into account. Conditions for the appearance of a
spin reorientation transition are given and analytic results for the width
of the canted phase and its shift in applied magnetic fields associated
with this transition are derived. 
\end{abstract}
\maketitle

\section{Introduction}

Experimentally it became possible in recent years to grow epitaxial thin
films of ferromagnetic materials on non-magnetic substrates with a very
high quality. This offers the possibility to stabilize crystallographic
structures which are not present in nature, and which may exhibit new properties
of high technological impact. To understand the magnetic structure of these
systems is a challenging problem both experimentally and theoretically.

Generally speaking, for not too thin films the magnetization is in-plane
due to the dipole interaction (shape anisotropy). However, in very thin
films this may change due to the increasing importance of surface effects.
Indeed, at surfaces due to the broken symmetry uniaxial anisotropy energies
arise which in generally are much higher than in the bulk. These anisotropy
energies may favor a perpendicular orientation of magnetization \cite{Neel54}.
Additionally in the inner layers of the film due to strain induced distortion
bulk anisotropy energies may appear absent or very small in the ideal crystal.
As a consequence in these films a reorientation of the spontaneous magnetization
is observed either as function of film thickness or as a function of temperature.
This spin reorientation transition has been discussed extensively in the
past \cite{Allen94,Pappas90,Qiu93,Schulz94}.

Phenomenologically in order to describe the magnetic properties, anisotropy
coefficients \( K_{n} \) compatible with the underlying symmetry of the
film are introduced which are supposed to arise from an expansion of the
energy (or the free energy at finite temperatures) in terms of the orientation
of the magnetization vector relative to the film. These coefficients are
then studied experimentally (for a review see Ref.~\onlinecite{Gradmann93}).
In ferromagnetic resonance (FMR) experiments, for instance, these coefficients
directly enter the resonance frequency (for references see for instance
Ref.~\onlinecite{Bab96,Farle98}). 

Theoretically, it has been shown that the anisotropy coefficients \( K_{n}(T) \),
which are in general temperature dependent, can be calculated numerically
at finite temperatures within mean field theory, starting from a Hamiltonian
with microscopic constant anisotropy parameters \cite{Hucht97}. Furthermore,
the temperature dependence of the lowest order anisotropy \( K(T) \) was
determined analytically using a combination of mean field theory and first
order perturbation theory \cite{Jensen96,Hucht97}. In other approaches
the magnetization of the film is calculated directly within mean field
and spin wave theory \cite{Moschel94,Moschel95,Froebrich00,Froebrich00a}
or with full numerical calculations like Monte Carlo simulations \cite{Hucht95, Hucht96}. 

In the present paper we describe the ferromagnetic film within a classical
local spin model with dipolar interaction and uniaxial anisotropy. We will
concentrate on ground-state properties of thin films in order to clarify
the discussion and to eliminate all uncertainties connected with finite
temperature calculations. A major goal of the present study is the calculation
of the anisotropy coefficients at zero temperature from the parameters
of an underlying Hamiltonian. The important point is that even in this
situation the dependence of these coefficients on the microscopic parameters
is far from being trivial due to non-collinear magnetic states in the thin
film. It is the purpose of this paper to elucidate this behavior.

\section{The model}

The calculations of the ground-state properties of ultra-thin ferromagnetic
films are done within the framework of a classical ferromagnetic Heisenberg
model consisting of \( L \) two-dimensional layers with the \( \vec{z} \)-direction
normal to the film. The Hamiltonian reads \begin{eqnarray}
\mathcal{H} & = & -\frac{J}{2}\sum _{\langle ij\rangle }\vec{s}_{i}\cdot \vec{s}_{j}+\frac{\omega }{2}\sum _{i\neq j}\frac{\vec{s}_{i}\cdot \vec{s}_{j}}{r_{ij}^{3}}-\frac{3(\vec{s}_{i}\cdot \vec{r}_{ij})(\vec{r}_{ij}\cdot \vec{s}_{j})}{r_{ij}^{5}}\nonumber \\
 &  & {}-\sum _{i}D^{(2)}_{\lambda _{i}}(s_{i}^{z})^{2}-\sum _{i}D^{(4)}_{\lambda _{i}}(s_{i}^{z})^{4}-\sum _{i}\vec{B}\cdot \vec{s}_{i},\label{e:H} 
\end{eqnarray}
 where \( \vec{s}_{i}=(s_{i}^{x},s_{i}^{y},s_{i}^{z}) \) are spin vectors
of unit length at position \( \vec{r}_{i}=(r_{i}^{x},r_{i}^{y},r_{i}^{z}) \)
in layer \( \lambda _{i} \) and \( \vec{r}_{ij}=\vec{r}_{i}-\vec{r}_{j} \).
The positions \( \vec{r}_{i} \) are normalized such that nearest neighbors
obey \( r_{\langle ij\rangle }=1 \). \( J \) is the nearest-neighbor
exchange coupling constant, \( D^{(2)}_{\lambda _{i}} \) and \( D^{(4)}_{\lambda _{i}} \)
are the local uniaxial anisotropies of second and fourth order, respectively,
\( \vec{B} \) denotes the external magnetic field with the effective magnetic
moment \( \mu  \) of the spins incorporated, and \( \omega =\mu _{0}\mu ^{2}/4\pi a^{3} \)
is the strength of the long range dipole interaction on a lattice with
lattice constant \( a \) (\( \mu _{0} \) is the magnetic permeability). 

To calculate the ground-state energy per spin we assume translational invariance
of the spin structure parallel to the film. This assumption is not correct
rigorously since it can be shown that for a perpendicular oriented magnetization,
for instance, a state with striped domains is energetically slightly more
favorable. However the corresponding energy difference for ultra-thin films
is of order \( e^{-J/2\omega } \) and therefore negligible for realistic
parameters of Fe- or Ni-films showing spin reorientation transitions \cite{Hucht96}.

Assuming translational invariance in the \( xy \)-plane the summation
over all spins within a plane can be done exactly resulting in the energy
per surface spin \begin{eqnarray}
E(\vec{s}) & = & -\frac{J}{2}\sum _{\mu ,\nu =1}^{L}z_{|\mu -\nu |}\, \vec{s}_{\mu }\cdot \vec{s}_{\nu }\nonumber \label{e:EGS} \\
 &  & {}-\frac{\omega }{2}\sum _{\mu ,\nu =1}^{L}\Phi _{|\mu -\nu |}\, \vec{s}_{\mu }\cdot \left( \begin{array}{ccr}
\frac{1}{2} & 0 & 0\\
0 & \frac{1}{2} & 0\\
0 & 0 & -1
\end{array}\right) \cdot \vec{s}_{\nu }\nonumber \\
 &  & {}-\sum _{\mu =1}^{L}D^{(2)}_{\mu }(s_{\mu }^{z})^{2}+D^{(4)}_{\mu }(s_{\mu }^{z})^{4}+\vec{B}\cdot \vec{s}_{\mu }\label{e:E(vecs)} 
\end{eqnarray}
 with \( \vec{s}=(\vec{s}_{1},\ldots ,\vec{s}_{L}) \). The quantities
\( \mu  \) and \( \nu  \) denote layer indices, \( z_{|\mu -\nu |} \)
is the number of nearest neighbors between layer \( \mu  \) and \( \nu  \),
and \( \Phi _{|\mu -\nu |} \) are constants arising from a partial summation
of the dipole interaction. The quantities \( \Phi _{\delta } \) have been
calculated previously \cite{Hucht98a,DrHucht} and they are listed together
with \( z_{\delta } \) in Table ~\ref{t:z+Phi}.

\begin{table*}
\def\arraystretch{1.4}\tabcolsep10pt \centering 

\begin{tabular}{r|c|c|c|c|c|l}
 lattice&
 \( z_{0} \)&
 \( z_{1} \)&
 \( z_{\delta >1} \)&
 \( \Phi _{0} \)&
 \( \Phi _{1} \)&
\multicolumn{1}{c}{\( \Phi _{\delta >1} \)}\\
\hline
sc(001)&
 \( 4 \)&
 \( 1 \)&
 \( 0 \)&
 \( 9.0336 \)&
 \( -0.3275 \)&
 \( \sim -16\pi ^{2}e^{-2\pi \delta } \)\\
 fcc(001)&
 \( 4 \)&
 \( 4 \)&
 \( 0 \)&
 \( 9.0336 \)&
 \( 1.4294 \)&
 \( \sim \mp 16\pi ^{2}e^{-\sqrt{2}\pi \delta } \)\\
 bcc(001)&
 \( 0 \)&
 \( 4 \)&
 \( 0 \)&
 \( 5.8675 \)&
 \( 2.7126 \)&
 \( \sim \mp 6\sqrt{3}\pi ^{2}e^{-\pi \delta } \)\\
\end{tabular}

\caption{Number of nearest neighbors \protect\( z_{\delta }\protect \) and dipole
sums \protect\( \Phi _{\delta }\protect \) for different lattice types.
\protect\( \delta \protect \) is the distance between layers. \label{t:z+Phi}}
\end{table*}

With an external magnetic field \( \vec{B}=(0,B_{\para },B_{\perp }) \)
in the \( yz \)-plane, all spins \( \vec{s}_{\mu } \) are confined to
this plane. They therefore can be expressed by their azimuthal angle \( \vartheta _{\mu } \),
\( \vec{s}_{\mu }=(0,\sin \vartheta _{\mu },\cos \vartheta _{\mu }) \).
Eq.~(\ref{e:EGS}) thus can be rewritten as \begin{eqnarray}
E(\vec{\vartheta }) & = & -\frac{1}{2}\sum _{\mu ,\nu =1}^{L}\left[ \left( Jz_{|\mu -\nu |}-\frac{\omega }{4}\Phi _{|\mu -\nu |}\right) \cos (\vartheta _{\mu }-\vartheta _{\nu })\vphantom {\frac{3\omega }{4}}\right. \nonumber \\
 &  & \qquad \qquad \left. {}-\frac{3\omega }{4}\Phi _{|\mu -\nu |}\cos (\vartheta _{\mu }+\vartheta _{\nu })\right] \nonumber \\
 &  & {}-\sum _{\mu =1}^{L}\left[ D^{(2)}_{\mu }\cos ^{2}\vartheta _{\mu }+D^{(4)}_{\mu }\cos ^{4}\vartheta _{\mu }\right. \nonumber \\
 &  & \qquad \qquad \left. {}+B_{\para }\sin \vartheta _{\mu }+B_{\perp }\cos \vartheta _{\mu }\vphantom {D^{(4)}_{\mu }}\right] \label{e:E(vecth)} 
\end{eqnarray}
 with \( \vec{\vartheta }=(\vartheta _{1},\ldots ,\vartheta _{L}) \).
The ground state is obtained by minimizing the energy \( E(\vec{\vartheta }) \)
with respect to \( \vec{\vartheta } \). In zero external field two stationary
points of the energy given in Eq.~(\ref{e:E(vecth)}) are easily identified
to be given by \( \vec{\vartheta }^{\perp }=(0,\ldots ,0) \) and \( \vec{\vartheta }^{\para }=(\frac{\pi }{2},\ldots ,\frac{\pi }{2}) \),
respectively. We define a total anisotropy per surface spin in zero field
\( K \) by the corresponding energy difference, \( K=E(\vec{\vartheta }^{\para })-E(\vec{\vartheta }^{\perp }) \).
This quantity is given by \begin{equation}
\label{e:K}
K=\sum _{\mu =1}^{L}\left( D^{(2)}_{\mu }+D^{(4)}_{\mu }\right) -\frac{3\omega }{4}\sum _{\mu ,\nu =1}^{L}\Phi _{|\mu -\nu |}.
\end{equation}
 The first term is the sum of the anisotropy constants of second and fourth
order while the second term is due to the dipole interaction. Note that
this dipole term is identical to the dipole anisotropy per unit area \( \frac{L}{2}\mu _{0}m^{2} \)
calculated within continuum theory, but with additional surface correction,
as \begin{equation}
\label{e:DipolKonti}
\frac{3\omega }{4}\sum _{\mu ,\nu =1}^{L}\Phi _{|\mu -\nu |}=\frac{L}{2}\mu _{0}m^{2}-\frac{3\omega }{2}\Phi _{1}+\mathcal{O}(\Phi _{2}).
\end{equation}

For \( K>0 \) a perpendicular magnetization is more favorable than an
in-plane magnetization and vice versa. However, in certain parameter intervals
additional stationary points appear which may lead to an even lower energy
resulting in a canted spin structure. This will be discussed in detail
in Section~\ref{s:RT}.

In general the minimization of Eq.~(\ref{e:E(vecth)}) has to be done
numerically. For realistic parameters appearing for instance for Fe- or
Ni films, however, the exchange interaction is by far the largest term
in the Hamiltonian leading to a nearly collinear spin structure. In this
situation the anisotropy terms can be treated as small perturbation and
as a consequence the minimization can be done to a large extend analytically.

\section{Perturbation calculation}

We define an averaged angle, \( \theta =\frac{1}{L}\sum _{\nu =1}^{L}\vartheta _{\nu } \)
and deviations from it, \( \epsilon _{\nu } \), so that \( \vartheta _{\nu }=\theta +\epsilon _{\nu } \)
and \( \sum _{\nu =1}^{L}\epsilon _{\nu }=0 \). Finite \( \epsilon _{\nu } \)
appear due to the various anisotropy terms and they are therefore small
for anisotropy terms (including the external magnetic field) which are
small compared to the exchange energy. This will be assumed in the following.
Under these circumstances a perturbative treatment is possible. We decompose
the energy Eq.~(\ref{e:E(vecth)}) into two parts, \begin{equation}
\label{e:E(vecth)2}
E(\vec{\vartheta })=E^{(0)}(\theta )+\delta E(\theta ,\vec{\epsilon })
\end{equation}
 with \( \delta E(\theta ,\vec{0})=0 \) and

\begin{eqnarray}
E^{(0)}(\theta ) & = & -\frac{J}{2}\sum _{\mu ,\nu =1}^{L}z_{|\mu -\nu |}+\frac{3\omega }{8}\cos (2\theta )\sum _{\mu ,\nu =1}^{L}\Phi _{|\mu -\nu |}\nonumber \\
 &  & {}-\cos ^{2}\theta \sum _{\mu =1}^{L}D^{(2)}_{\mu }-\cos ^{4}\theta \sum _{\mu =1}^{L}D^{(4)}_{\mu }\nonumber \\
 &  & {}-L(B_{\para }\sin \theta +B_{\perp }\cos \theta )\vphantom {\sum _{\mu =1}^{L}}\label{e:E0(th)} 
\end{eqnarray}
 An expansion of \( \delta E(\theta ,\vec{\epsilon }) \) in terms of \( \vec{\epsilon } \)
then gives \begin{equation}
\label{e:dE(th,eps)}
\delta E(\theta ,\vec{\epsilon })=\vec{a}(\theta )\cdot \vec{\epsilon }+\frac{1}{2}\, \vec{\epsilon }\cdot \mat{C}\cdot \vec{\epsilon }+\mathcal{O}(\vec{\epsilon })^{3}
\end{equation}
 where we have introduced an obvious matrix notation. The gradient \begin{equation}
\label{e:a(th)}
\vec{a}(\theta )=\left. \frac{\partial }{\partial \vec{\epsilon }}\, \delta E(\theta ,\vec{\epsilon })\right| _{\vec{\epsilon }\, =\, \vec{0}}
\end{equation}
 is given by \begin{equation}
\label{e:a(th)2}
\vec{a}(\theta )=\vec{A}(\theta )\sin (2\theta )
\end{equation}
 with \begin{equation}
\label{e:Al(th)}
A_{\lambda }(\theta )=D^{(2)}_{\lambda }+2D^{(4)}_{\lambda }\cos ^{2}\theta -\frac{3\omega }{4}\sum _{\mu =1}^{L}\Phi _{|\lambda -\mu |}.
\end{equation}
 Thus, to lowest order the anisotropy terms are linear in \( \vec{\epsilon } \)
while the exchange term expressed in Eq.~(\ref{e:dE(th,eps)}) by the
matrix \( \mat{C} \) with matrix elements \begin{equation}
\label{e:Cmn}
C_{\mu \nu }=-Jz_{|\mu -\nu |}+\delta _{\mu \nu }\sum _{\lambda =1}^{L}Jz_{|\mu -\lambda |}
\end{equation}
 is quadratic in \( \vec{\epsilon } \). 

The minimum of \( \delta E(\theta ,\vec{\epsilon }) \) appears for \( \epsilon _{\nu } \)
of the order of the anisotropy terms showing that the truncated Eq.~(\ref{e:dE(th,eps)})
gives the correct energy up to second order in \( \vec{\epsilon } \).
Note that up to this order the Zeeman term enters only Eq.~(\ref{e:E0(th)}).
Therefore, at this level of truncation \( \theta  \) agrees with the azimuthal
angel of the averaged magnetization.

It can be easily seen from the definition Eq.~(\ref{e:Cmn}) that \( \vec{e}_{0}=(1,\ldots ,1) \)
is an eigenvector of \( \mat{C} \) with eigenvalue zero. With this vector
it is convenient to rewrite the constrain \( \sum _{\nu =1}^{L}\epsilon _{\nu }=0 \)
as a scalar product, \( \vec{e}_{0}\cdot \vec{\epsilon }=0 \). This notation
will be used in the following.

The minimalization of the energy is done in two steps. First we keep \( \theta  \)
fixed and minimize with respect to \( \epsilon _{\nu } \) under the constraint
\( \vec{e}_{0}\cdot \vec{\epsilon }=0 \). The corresponding energy at
the minimum, \( E(\theta ) \), is accessible for instance by varying the
external magnetic field and it is precisely this quantity which for instance
is needed to calculate the FMR signal. Finally the ground state energy
is obtained by minimizing \( E(\theta ) \) with respect to \( \theta  \).

The variation with respect to \( \epsilon _{\nu } \) is achieved by introducing
the function \begin{equation}
\label{e:Psi(th,eps)}
\Psi (\theta ,\vec{\epsilon })=\vec{a}(\theta )\cdot \vec{\epsilon }+\frac{1}{2}\, \vec{\epsilon }\cdot \mat{C}\cdot \vec{\epsilon }+\lambda \, \vec{e}_{0}\cdot \vec{\epsilon }
\end{equation}
 where \( \lambda  \) denotes a Lagrangian multiplier. Stationarity of
\( \Psi (\theta ,\vec{\epsilon }) \) gives \begin{equation}
\label{e:stat}
\mat{C}\cdot \vec{\epsilon }+\vec{a}(\theta )+\lambda \, \vec{e}_{0}=\vec{0}.
\end{equation}
 Taking the scalar product with \( \vec{e}_{0} \) and noting that \( \vec{e}_{0}\cdot \mat{C}=\vec{0} \)
the multiplier \( \lambda  \) is obtained as \begin{equation}
\label{e:lambda}
\lambda =-\frac{1}{L}\, \vec{e}_{0}\cdot \vec{a}(\theta ).
\end{equation}
 Thus \( \vec{\epsilon } \) is determined from \begin{equation}
\label{e:eps1}
\mat{C}\cdot \vec{\epsilon }+\left( \mat{1}-\frac{1}{L}\mat{E}\right) \cdot \vec{a}(\theta )=\vec{0}
\end{equation}
 with identity matrix \( \mat{1} \) and a matrix \( \mat{E} \) with \( E_{\mu \nu }=1 \)
for all matrix elements. To solve this equation for \( \vec{\epsilon } \)
we introduce the pseudo-inverse \( \mat{C}^{\dagger } \) of the matrix
\( \mat{C} \), which in our case fulfills \begin{equation}
\label{e:Gdef}
\mat{C}\cdot \mat{C}^{\dagger }=\mat{1}-\frac{1}{L}\mat{E}.
\end{equation}
 The matrix \( \mat{C}^{\dagger } \) is uniquely defined if one requires
that it is a symmetric matrix with eigenvector \( \vec{e}_{0} \) and corresponding
eigenvalue zero. The matrix elements of \( \mat{C}^{\dagger } \) are explicitly
given by \cite{DrHucht} \begin{eqnarray}
C^{\dagger }_{\mu \nu } & = & \frac{1}{2LJz_{1}}\left[ \frac{L^{2}-1}{6}-L|\mu -\nu |\right. \nonumber \\
 &  & {}+\left. \left( \mu -\frac{L+1}{2}\right) ^{2}+\left( \nu -\frac{L+1}{2}\right) ^{2}\right] .\label{e:Gmn} 
\end{eqnarray}
 It is easy to see that with the help of this matrix Eq.~(\ref{e:eps1})
can be rewritten as \begin{equation}
\label{e:eps2}
\mat{C}\cdot \left( \vec{\epsilon }+\mat{C}^{\dagger }\cdot \vec{a}(\theta )\right) =\vec{0}.
\end{equation}
 Since \( \vec{e}_{0} \) is the only eigenvector of \( \mat{C} \) with
eigenvalue zero the term in brackets has to be parallel to \( \vec{e}_{0} \).
Multiplying this term by \( \vec{e}_{0} \) and using \( \vec{e}_{0}\cdot \vec{\epsilon }=0 \)
and \( \vec{e}_{0}\cdot \mat{C}^{\dagger }=\vec{0} \) it follows \begin{equation}
\label{e:eps3}
\vec{\epsilon }=-\mat{C}^{\dagger }\cdot \vec{a}(\theta ).
\end{equation}
 Inserting into Eq.~(\ref{e:Psi(th,eps)}) we get the final result \bS
\label{e:E(th)} \begin{eqnarray}
E(\theta ) & = & E^{(0)}(\theta )+\delta E(\theta )\\
\delta E(\theta ) & = & -\frac{1}{2}\, \vec{a}(\theta )\cdot \mat{C}^{\dagger }\cdot \vec{a}(\theta )+\mathcal{O}(\vec{\epsilon })^{3},\label{e:dE(th)} 
\end{eqnarray}
\eS where we used the general property \( \mat{C}^{\dagger }=\mat{C}^{\dagger }\cdot \mat{C}\cdot \mat{C}^{\dagger } \)
of the pseudo inverse. The ground state energy is obtained by minimizing
\( E(\theta ) \) with respect to \( \theta  \).

Eq.~(\ref{e:E(th)}) is the main result of this work, giving a general
expression for the ground state energy of a thin magnetic film in second
order perturbation theory. The influence of a non-collinear spin structure
on the ground state energy will be discussed in the following.

\section{Results}

In the following we drop terms of order \( \mathcal{O}(\vec{\epsilon })^{3} \)
in \( E(\theta ) \) and we specialize to a special case in order to obtain
analytic results. We neglect the exponentially small effective dipole interactions
between layers with distance larger that one, i.e. \( \Phi _{\delta >1}=0 \),
and we assume that the anisotropies \( D_{\lambda }^{(n)} \) which enter
the Hamiltonian Eq.~(\ref{e:H}) are constant within the thin film but
may deviate from its constant value at the surface (\( \lambda =1 \))
and at the interface to the substrate (\( \lambda =L \)), i.e. \begin{eqnarray}
D_{\lambda }^{(n)} & = & D_{\mathrm{v}}^{(n)}+\delta _{\lambda ,1}D_{\mathrm{s}}^{(n)}+\delta _{\lambda ,L}D_{\mathrm{i}}^{(n)}\label{e:Dsidef} \\
A_{\lambda }(\theta ) & = & A_{\mathrm{v}}(\theta )+\delta _{\lambda ,1}A_{\mathrm{s}}(\theta )+\delta _{\lambda ,L}A_{\mathrm{i}}(\theta )\label{e:Asi(th)def} 
\end{eqnarray}
 with\begin{equation}
\label{e:Asi(th)}
A_{\mathrm{s},\mathrm{i}}(\theta )=D^{(2)}_{\mathrm{s},\mathrm{i}}+2D^{(4)}_{\mathrm{s},\mathrm{i}}\cos ^{2}\theta +\frac{3\omega }{4}\Phi _{1}
\end{equation}
 It is easy to see that \begin{eqnarray}
\vec{A}(\theta )\cdot \mat{C}^{\dagger }\cdot \vec{A}(\theta ) & = & C^{\dagger }_{1,1}(A_{\mathrm{s}}^{2}(\theta )+A_{\mathrm{i}}^{2}(\theta ))\nonumber \\
 &  & {}+2C^{\dagger }_{1,L}A_{\mathrm{s}}(\theta )A_{\mathrm{i}}(\theta )\label{e:ACA} 
\end{eqnarray}
 since \( \mat{C}^{\dagger }\cdot \vec{e}_{0}=\vec{0} \) and \( C^{\dagger }_{1,1}=C^{\dagger }_{L,L} \).
Then the second order correction calculated in the previous section (Eq.~(\ref{e:dE(th)}))
can be written as \begin{equation}
\label{e:dE(th)2}
\delta E(\theta )=\Delta (\theta ,L)\sin ^{2}(2\theta )
\end{equation}
 with \begin{eqnarray}
\Delta (\theta ,L) & = & -\frac{L-1}{8Jz_{1}}\left[ \frac{L-2}{3L}\left( A_{\mathrm{s}}(\theta )+A_{\mathrm{i}}(\theta )\right) ^{2}\right. \nonumber \\
 &  & \qquad \qquad \left. {}+\left( A_{\mathrm{s}}(\theta )-A_{\mathrm{i}}(\theta )\right) ^{2}\right] .\label{e:D(th,L)} 
\end{eqnarray}
 Note that from now on \( L \) can be considered as continuous parameter
and all quantities are explicitly \( L \)--dependent. Inserting \( \vec{a}(\theta ) \)
and \( E^{(0)}(\theta ) \) into Eq.~(\ref{e:E(th)}) and introducing
the quantities \bS\begin{eqnarray}
K_{0}(L) & = & -\frac{J}{2}(zL-2z_{1})-\frac{\omega }{2}\left( 2\pi L-\frac{3}{2}\Phi _{1}\right) \label{e:K0(L)} \\
K_{2}(L) & = & LD^{(2)}_{\mathrm{v}}+D^{(2)}_{\mathrm{s}}+D^{(2)}_{\mathrm{i}}\label{e:K2(L)} \\
 &  & {}-\omega \left( 2\pi L-\frac{3}{2}\Phi _{1}\right) \nonumber \\
K_{4}(L) & = & LD^{(4)}_{\mathrm{v}}+D^{(4)}_{\mathrm{s}}+D^{(4)}_{\mathrm{i}}\label{e:K4(L)} 
\end{eqnarray}
 \eS we can finally write for the energy per surface spin \begin{eqnarray}
E(\theta ,L) & = & K_{0}(L)+\Delta (\theta ,L)\, \sin ^{2}(2\theta )\nonumber \\
 &  & {}-K_{2}(L)\cos ^{2}\theta -K_{4}(L)\cos ^{4}\theta \nonumber \\
 &  & {}-L(B_{\para }\sin \theta +B_{\perp }\cos \theta )\label{e:E(th,L)} 
\end{eqnarray}
 Note that the total anisotropy energy \( K \) introduced in Eq.~(\ref{e:K})
fulfills \[
K(L)=K_{2}(L)+K_{4}(L),\]
 as \( \delta E(\theta ) \) vanishes at the collinear stationary points
\( \vec{\vartheta }^{\para } \) and \( \vec{\vartheta }^{\perp } \),
respectively. \( K_{2}(L) \) and \( K_{4}(L) \) contain the microscopic
anisotropy parameters and the dipole terms of the film averaged over the
different layers.

It is easy to see that an equation for \( E(\theta ,L) \) in the form
given by Eq.~(\ref{e:E(th,L)}) often introduced phenomenologically \cite{Farle98},
but without the \( \Delta  \)--term, is obtained if one assumes that all
spins in the film are strictly parallel. The important point to note here,
however, is the fact that an additional anisotropy energy \( \Delta (\theta ,L) \)
enters Eq.~(\ref{e:E(th,L)}) which is connected to non-collinear spin
structures originated by inhomogeneities in the magnetic film. Indeed,
this quantity only vanishes in the homogeneous case \( A_{\mathrm{s}}=A_{\mathrm{i}}=0 \).
However, for a magnetic thin film the amplitudes \( A_{\lambda } \) in
general are not constant. Even if the microscopic anisotropy constants
\( D^{(n)}_{\lambda } \) are homogeneous (which is unlikely to occur for
a realistic film) this is not the case for the dipole term.

To discuss the implications of this additional anisotropy term \( \Delta (\theta ,L \))
we first consider the case that there is no microscopic uniaxial anisotropy
of fourth order, \( D^{(4)}_{\lambda }=0 \). In this case \( \Delta (\theta ,L)=\Delta (L) \)
is independent of \( \theta  \). Thus for an inhomogeneous distribution
of amplitudes \( A_{\lambda } \), an effective anisotropy term of fourth
order in \( \cos \theta  \) is generated although there is no corresponding
anisotropy term of this order in the Hamiltonian.

If there exists a microscopic anisotropy term of fourth order the situation
is more complicated: \( \Delta  \) becomes \( \theta  \)-dependent meaning
that higher order anisotropy term of up to eights order are generated in
\( E(\theta ,L) \).

Finally we mention that the quantity \( \Delta (\theta ,L) \) can be further
simplified in two common special cases: In the case of a symmetric film
\( D^{(n)}_{\mathrm{i}}=D^{(n)}_{\mathrm{s}} \) we get \( A_{\mathrm{s}}(\theta )=A_{\mathrm{i}}(\theta ) \)
and therefore \bS \begin{equation}
\label{e:Dis(th,L)}
\Delta _{\mathrm{i}=\mathrm{s}}(\theta ,L)=-\frac{(L-1)(L-2)}{6LJz_{1}}A^{2}_{\mathrm{s}}(\theta ),
\end{equation}
 while for the case \( D^{(n)}_{\mathrm{i}}=0 \) and \( D^{(2)}_{\mathrm{s}}+D^{(4)}_{\mathrm{s}}\gg \frac{3\omega }{2}\Phi _{1} \)
we have \( A_{\mathrm{s}}(\theta )\gg A_{\mathrm{i}}(\theta ) \) and \begin{equation}
\label{e:Di0(th,L)}
\Delta _{\mathrm{i}=0}(\theta ,L)=-\frac{(L-1)(L-\frac{1}{2})}{6LJz_{1}}A_{\mathrm{s}}^{2}(\theta ).
\end{equation}
\eS

As an important application of these results we will study spin reorientation
transitions in the next section.

\section{Spin Reorientation transition \label{s:RT}}

The direction of the magnetization in the ground state for a given thickness
\( L \) is obtained by minimizing \( E(\theta ,L) \) (Eq.~(\ref{e:E(th,L)})).
If the total anisotropy energy \( K(L) \) (Eq.~(\ref{e:K(L)})) changes
sign as function of \( L \), a spin reorientation transition takes place
in which the direction of the magnetization changes either continuously
or discontinuously depending on the specific form of \( E(\theta ,L) \).
In the first case a so-called canted phase appears. Analytic results for
the width and the position of this phase will be derived in this chapter.

We decompose \( K(L) \) in volume and surface part the usual way \cite{Gradmann86}
to get \begin{equation}
\label{e:K(L)}
K(L)=LK_{\mathrm{v}}+K_{\mathrm{s}}+K_{\mathrm{i}}
\end{equation}
 with \bS \begin{eqnarray}
K_{\mathrm{v}} & = & D^{(2)}_{\mathrm{v}}+D^{(4)}_{\mathrm{v}}-2\pi \omega \label{e:Kv} \\
K_{\mathrm{s},\mathrm{i}} & = & D^{(2)}_{\mathrm{s},\mathrm{i}}+D^{(4)}_{\mathrm{s},\mathrm{i}}+\frac{3\omega }{4}\Phi _{1}\label{e:Ksi} 
\end{eqnarray}
 \eS Note that \( A_{\mathrm{s},\mathrm{i}}(\theta ) \) from Eq.~(\ref{e:Asi(th)})
can be written as \begin{equation}
\label{e:Asi(th)2}
A_{\mathrm{s},\mathrm{i}}(\theta )=K_{\mathrm{s},\mathrm{i}}+D^{(4)}_{\mathrm{s},\mathrm{i}}\cos 2\theta .
\end{equation}
 A spin reorientation transition occurs if the total anisotropy energy
\( K(L) \) passes through zero as function of \( L \). If \( K_{\mathrm{s}}+K_{\mathrm{i}}>0 \)
this happens for sufficiently large dipole interaction with increasing
\( L \), as then \( K_{\mathrm{v}}<0 \). The corresponding transition
is from perpendicular magnetization at small \( L \) to an in-plane magnetization
for large \( L \) possibly with a canted magnetization in between. This
type of transition occurs for Fe-films. The opposite scenario can occur
for negative \( K_{\mathrm{s}}+K_{\mathrm{i}} \) if a positive volume
anisotropy \( K_{\mathrm{v}}>0 \) is present as observed in Ni-films.
Thus, to lowest order the critical thickness is explicitly given by \( K(L_{\mathrm{r}})=0 \),
leading to \begin{equation}
\label{e:Lr}
L_{\mathrm{r}}=-\frac{K_{\mathrm{s}}+K_{\mathrm{i}}}{\mathrm{K}_{\mathrm{v}}}.
\end{equation}
 For Fe/Ag(100) films \( D^{(2)}_{\mathrm{s}}+D^{(2)}_{\mathrm{i}}\approx 37\omega  \).
In this case the other quantities \( D^{(4)}_{\mathrm{v}} \), \( D^{(2)}_{\mathrm{v}} \)
and \( \Phi _{1} \) are negligible and we get \( L_{\mathrm{r}}\approx 5.5 \)
in good agreement with numerical calculations \cite{DrHucht}.

For \( L \) in the vicinity of \( L_{\mathrm{r}} \) the minimum of \( E(\theta ) \)
may occur at a finite \( \theta  \), i.e. a canted phase occurs. To deduce
the limits of stability of the two phases for which \( \theta =0 \) and
\( \theta =\frac{\pi }{2} \), respectively, we expand Eq.~(\ref{e:E(th,L)})
around these angles. From the sign of the corresponding expansion coefficient
it follows that in general there are two transitions of second order at
thicknesses \( L_{\mathrm{r}}^{\para } \) and \( L_{\mathrm{r}}^{\perp } \),
respectively. The phase with \( \theta =0 \) becomes unstable at \( L_{\mathrm{r}}^{\para } \)
where \bS \begin{equation}
\label{e:Lrp}
K(L_{\mathrm{r}}^{\para })+K_{4}(L_{\mathrm{r}}^{\para })+4\Delta (0,L_{\mathrm{r}}^{\para })=0
\end{equation}
 at this point. With increasing thickness the parallel phase with \( \theta =\frac{\pi }{2} \)
becomes stable at \( L_{\mathrm{r}}^{\perp } \) where \begin{equation}
\label{e:Lrs}
\textstyle K(L_{\mathrm{r}}^{\perp })-K_{4}(L_{\mathrm{r}}^{\perp })-4\Delta (\frac{\pi }{2},L_{\mathrm{r}}^{\perp })=0.
\end{equation}
 \eS

For \( K_{4}(L_{\mathrm{r}})+4\Delta (L_{\mathrm{r}})=0 \) both transitions
coincide resulting in a jump from \( \theta =0 \) to \( \theta =\frac{\pi }{2} \)
at \( L_{\mathrm{r}} \). This is always the case for \( L=1 \) and in
the symmetric case also for \( L=2 \) provided \( D_{\lambda }^{(4)} \)
vanishes. Otherwise a canted phase (\( K_{4}+4\Delta >0 \)) or a region
with hysteresis (\( K_{4}+4\Delta <0 \)) appears as described below. Note
that in the phases \( \theta =0 \) and \( \theta =\frac{\pi }{2} \),
respectively, \( \vec{\epsilon } \) vanishes according to Eqs.~(\ref{e:a(th)2},
\ref{e:eps3}) showing that in these phases all spins are strictly parallel.
This is not the case in the canted phase. Note also that for finite magnetic
fields which are neither perpendicular nor parallel to the film minimalization
of Eq.~(\ref{e:E(th,L)}) leads to a \( \theta  \) between zero and \( \frac{\pi }{2} \)
and therefore to a noncollinear spin structure.

The difference of the thicknesses at which the two collinear phases become
instable defines the width \( \delta L_{\mathrm{r}}=L_{\mathrm{r}}^{\perp }-L_{\mathrm{r}}^{\para } \)
of the canted region which can be expressed as\begin{equation}
\label{e:dLr}
\frac{\delta L_{\mathrm{r}}}{L_{\mathrm{r}}}=-\frac{2K_{4}(L_{\mathrm{r}})+4(\Delta (0,L_{\mathrm{r}})+\Delta (\frac{\pi }{2},L_{\mathrm{r}}))}{K_{\mathrm{s}}+K_{\mathrm{i}}}
\end{equation}
 Thus the fourth order anisotropy energy \( D_{\lambda }^{(4)} \) increases
the width of the canted phase but even without such a term a canted region
can be observed due to the effective anisotropy \( \Delta (\theta ,L_{\mathrm{r}}) \).
If the numerator of the right hand side of Eq.~(\ref{e:dLr}) is positive,
a canted phase occur, while for negative numerator we find a discontinuous
transition with hysteresis.

A similar calculation can be done in finite magnetic fields. If the field
is orientated perpendicular to the film the thickness at which the phase
\( \theta =0 \) becomes instable is shifted by \bS \begin{equation}
\label{e:B1}
\frac{\delta L^{\para }_{\mathrm{r}}}{L_{\mathrm{r}}}=-\frac{B_{\perp }}{2\mathrm{K}_{\mathrm{v}}}.
\end{equation}
 while for fields parallel to the film the corresponding shift is given
by \begin{equation}
\label{e:B2}
\frac{\delta L^{\perp }_{\mathrm{r}}}{L_{\mathrm{r}}}=\frac{B_{\para }}{2\mathrm{K}_{\mathrm{v}}}.
\end{equation}
\eS

A phase diagram for finite temperatures and field has been obtained within
mean field theory previously \cite{Hucht99a}. For small external fields
the shifts of the phase boundaries obtained are linear in the field similar
to the present situation.

\section{Conclusion}

Starting from a microscopic model the ground state energy of a thin ferromagnetic
film as function of the direction of the magnetization is calculated. Explicit
expressions for this energy are obtained which contain important anisotropy
contributions due to non-collinear spin structures in certain parameter
intervals. The microscopic parameters entering the Hamiltonian are not
in a simple way related to the ground state energy. This is important for
a comparison of measured and calculated anisotropy parameters. Our investigation
shows that in generally a canted phase is obtained and that the corresponding
transitions into this phase are of second order. Analytic expressions are
obtained for the width of the canted phase and its shift in external magnetic
fields.

\begin{acknowledgments}
This work was supported by the Deutsche Forschungsgemeinschaft through
Sonderforschungsbereich 491.

\bibliographystyle{}
\bibliography{Physik}
\end{acknowledgments}

\end{document}